\documentclass[usenatbib]{mn2e}
\usepackage{epsfig}
\def\nei{\hbox{Ne\,{\sc i}}}
\def\neii{\hbox{Ne\,{\sc ii}}}
\def\hi{\hbox{H\,{\sc i}}}

\newcommand{\teff}{\mbox{$T_{\mathrm{eff}}$}}   

\begin{document}

\title[Radiative accelerations on Ne in the atmospheres of late B
stars]{Radiative accelerations on Ne in the atmospheres of late B
stars} \author[J. Budaj, M.M. Dworetsky]
{J. Budaj$^{1,2}$\thanks{E-mail: budaj@ta3.sk; at UCL in 1999-2000} 
and M.M. Dworetsky$^{1}$\\
$^1$Dept. of Physics \& Astronomy, University College London,
Gower Street, London WCIE 6BT, United Kingdom\\
$^2$Astronomical Institute, Slovak Academy of Sciences,
059 60 Tatransk\'{a} Lomnica, The Slovak Republic}

\date{Accepted 2002 August 21.  Received 2002 August 9; in original form 
2002 April 26}

\maketitle
\begin{abstract}
Radiative accelerations on Ne are calculated for the atmospheres of
main sequence stars with $11\,000 \leq \teff \leq 15\,000$\,K.  This range
corresponds to that of the mercury-manganese (HgMn) stars.  
The calculations take into account neon fine structure as well as
shadowing of neon lines using the entire Kurucz line list,
bound-bound, bound-free and free-free opacity of H, He and C as well as
some NLTE effects.  
NLTE effects are found to modify the radiative acceleration by a factor of
order $10^{2}$ in the outer atmosphere and are crucial
for $dm<10^{-3}$ g\,cm$^{-2}$.
The dependence of the radiative accelerations on the
Ne abundance, effective temperature and gravity is studied. 
Radiative
accelerations are found to be well below the gravity in the entire range
of \teff\ and it is predicted that in stable atmospheres devoid of
disturbing motions, Ne should sink and be observed as underabundant.  
This agrees with recent observations of low Ne abundances in HgMn stars.
\end{abstract}


\section{Introduction}

Radiatively driven diffusion processes are generally accepted to be
responsible for chemical peculiarities observed among upper main sequence
CP stars \citep{michaud70}. The upward radiative acceleration and downward
gravity acting on different chemical species compete and drive their
microscopic diffusion, resulting in the stratification of these elements
in the stellar atmospheres or envelopes of these stars. If these
atmospheres are sufficiently stable, chemical inhomogeneities can
persevere, and are not wiped out by various mixing processes such as
convective zones, stellar winds or rotationally induced mixing. In the
particular case of HgMn star atmospheres, temperatures are high enough for
hydrogen to be largely ionized, thus removing its convection zone.  Also,
if the He ionization convection zone diminishes due to He settling in late
B stars rotating slower than {\it{c.}} 75 km\,s$^{-1}$
\citep{michaud82,cm88}, diffusion can operate very effectively in the
atmosphere. This is the case for HgMn stars as they are also known as slow
rotators. Such diffusion processes are generally time dependent problems,
but the characteristic time-scales in the atmosphere are much shorter than
in the envelope ($\leq 10^{3}$ yr) due to the considerable lower density,
smaller collision rates and higher diffusion velocities
({\it{c.}}\,1\,cm\,s$^{-1}$). This makes the time dependent calculations
extremely difficult.  So far, diffusion can be followed in time only in
the envelope, as shown in the case of neon by \citet{Seaton97,Seaton99};
\citet*{ldv98} and \citet*{lmr00}. These calculations suggest that
radiative
accelerations on Ne are much lower than gravity below the atmosphere and
Ne should sink and be observed as underabundant unless a weak stellar wind
transports upwards the Ne accumulated in deep layers.

Nevertheless, observations are restricted to the atmospheres. This
important stellar region bridges the effects in the envelopes with those
outside the stars, such as stellar winds, and one clearly needs also to
know the radiative accelerations here in order to compare the theory
with observations. The case of Ne is worth studying for two reasons: it is
an important element having very high standard abundance
\citep*{ag89,gns96,db00} of about A(Ne/H)=$1.23 \times 10^{-4}$, and
because one can expect similar effects as in the case of He. \nei\ (along
with He\,{\sc i}) has a very high ionization potential and first excited
level well above the ground state. \nei\ is thus the dominant ionization
state in the atmospheres of these stars and all \nei\ resonance lines are
in the Lyman continuum. This part of the spectrum is very sensitive to
non-LTE (NLTE) effects and the temperature structure of the atmosphere. It
is known
that there is a big difference in the temperature behaviour between LTE
and NLTE model atmospheres at small optical depths and that temperature
can rise outwards in the NLTE models. \nei\ is also known to have strong
NLTE effects on lines originating from excited levels in the visible part
of the spectrum even in late B-type main sequence stars
\citep{am73,sigut99,db00}. Thus, the main complication is clear in doing
the task.  One cannot rely upon the diffusion approximation when
calculating energy flux as this is a non-local quantity in the atmosphere,
in contrast to envelope calculations. One needs to solve the radiative
transfer as, e.g., in \citet*{hlh00}, \citet*{haa96} or \citet{hlhb02}, 
including NLTE effects if possible.

Unfortunately, observations of Ne in HgMn stars are very scarce and apart
from recent LTE Ne abundances in $\kappa$\,Cnc, HR7245 and HD29647 by 
\citet{ap00} and \citet{aswesef01}
the only study of Ne was undertaken recently by \citet{db00}. Their NLTE
analysis of 20 HgMn and 5 normal stars revealed that Ne is clearly
underabundant in HgMn stars and the deficit reaches its peak values in the
middle of the \teff\ domain of HgMn stars.

In this paper we deal with the radiative accelerations of neon in the
atmosphere.  This might enable us to model the vertical element
stratification in the future, e.g. considering the atmosphere as being in
a sequence of stationary states which are to be tailored to the much more
slowly varying envelope as proposed by \citet{bzzzk93}.  
Nevertheless, some conclusions can be drawn already from the radiative
accelerations alone. In the following, cgs units and LTE approximations
(in a NLTE atmosphere model) are used if not specified otherwise.

\section{Radiative acceleration}

\subsection{Theory}
\label{theory}

The physics and calculations of radiative accelerations in the envelopes
are clearly explained in \citet{Seaton97}, recent improvements are
described in \citet{glam95} and some theoretical aspects are discussed in
\citet{sa95}.  Our approach to the problem of calculating accelerations in
the atmospheres is described here. Radiative acceleration on the element
is due to the momentum transferred to the element ions from photons
absorbed via bound-bound, bound-free and free-free transitions.  
Bound-bound transitions are usually the most important. An expression for
the radiative acceleration on the ion $i$ through its spectral lines can
be derived from the radiative transfer equation. For a particular spectral
line in the case of complete redistribution we have:
\begin {equation}
dI_{\nu}=[n_{u,i}A_{ul}-(n_{l,i}B_{lu}-n_{u,i}B_{ul})I_{\nu}]h \nu 
\varphi_{lu}(\nu)ds
\label{e1}
\end {equation}
where the three terms describe spontaneous isotropic emission,
absorption, and
stimulated emission, respectively. Note that stimulated emission is in 
the direction of the photon absorbed. Here, and in the following,
$I_{\nu}$ is monochromatic intensity,
$F_{\nu}$ is monochromatic energy flux, 
$h \nu$ is the energy of the transition from the lower level $l$ to 
the upper level $u$,
$\varphi_{lu}(\nu,\tau)$ is the normalized Voigt profile,
$\tau$ is the Rosseland optical depth,
$s$ is the distance along the beam,
$n_{l,i}$ is the NLTE population of the $l$-th state of $i$-th ion,
$n_{i}$ is the total NLTE $i$-th ion population,
$b_{l,i}, b_{u,i}$ are b-factors defined as a ratio of NLTE to LTE level
population ($b_{l,i}\equiv n_{l,i}/n_{l,i}^{\ast}$),
$m$ is the mass of the $i$-th ion,
$c$ is the speed of light,
$k$ is the Boltzmann constant,
and $T$ is the temperature.
$A_{ul}, B_{ul}, B_{lu}$ are the Einstein coefficients where
$B_{lu}$ is related to the oscillator strength $f_{lu}$ and $B_{ul}$ by:
\begin {equation}
B_{lu}=\pi e^{2} f_{lu} (m_{\rm e}c h
\nu)^{-1},~~~g_{l,i}B_{lu}=g_{u,i}B_{ul}
\label{e2}
\end {equation}
where $e, m_{\rm e}$ are the electron charge and mass respectively
and $g_{l,i}, g_{u,i}$ are statistical weights of the lower and upper
levels.
Introducing the vertical geometrical depth $dr=\cos \theta ds$,
multiplying Eq. \ref{e1} by $\cos \theta d\omega$ and integrating
through all solid angles $d\omega$ we get for the radiation pressure 
gradient $dp_{\nu}^{R}/dr$:
\begin {equation}
c \frac{dp_{\nu}^{R}}{dr}=
-\left(1-\frac{n_{u,i}}{n_{l,i}}\frac{g_{l,i}}{g_{u,i}}
\right) n_{l,i} B_{lu} h \nu F_{\nu} \varphi_{lu}(\nu)
\label{e3}   
\end {equation}
If $p_{lu}$ is the momentum removed from the radiation field through 
one spectral line in unit volume and $a_{lu}$, $a_{i}$ are the corresponding
accelerations acquired by the ion $i$ then we get:
\begin {equation}
a_{i}(\tau) = \sum_{lu} a_{lu} =
\frac{1}{mn_{i}} \sum_{lu} \frac{dp_{lu}}{dt}~~~~~~\rm or 
\label{e4}
\end {equation}
\begin {equation}
a_{i}(\tau) =
- \frac{1}{m n_{i}} \sum_{lu} \int_{0}^{\infty} \frac{dp_{\nu}^{R}}{dr}d \nu
\label{e5}
\end {equation}
The sum runs through all $l \rightarrow u$ transitions in the ion.
Applying the Boltzmann formula to LTE populations, in this general NLTE
case, we finally have:
\[
a_{i}(\tau) = \sum_{lu} \frac {h \nu}{m c}
\left(1-\frac{b_{u,i}(\tau)}{b_{l,i}(\tau)} e^{-h
\nu/[kT(\tau)]}\right)B_{lu}
\frac{n_{l,i}(\tau)} {n_{i}(\tau)} 
\]
\begin {equation}
\int_{0}^{\infty} F_{\nu}(\tau) \varphi_{lu}(\nu, \tau) d \nu~~.
\label{e6}
\end {equation}
Note that, the particle flux of the
element associated with such definition of radiative acceleration would
be:
\footnote{One can define the radiative acceleration in some other way
as e.g. \citet{Seaton97}, corresponding to different expressions for the
particle flux.}
\begin {equation}
J =  \sum_{i} n_{i} v_{i} =
\sum_{i} n_{i} D_{i} \frac{m}{kT} a_{i}
\label{e7}  
\end {equation}
where $v_{i}, D_{i}$ are the $i$-th ion diffusion velocity and
diffusion coefficient, respectively.

The above expression for radiative acceleration includes the correction 
for stimulated emission which is often omitted by other authors, as
pointed
out by \citet{budaj94} and \citet{Seaton97}, but does not include the
redistribution effect pointed out by \citet{mm76} and generalized to
include gravity by \citet{av83}. It has been substantially 
revisited by \citet{glam95}.
The main idea of the redistribution effect is that,
after the photon absorption, ion $i$ is in the excited state $u$ and has
non-negligible probability  $(1-r_{u,i})$ of being ionized (mainly by 
collisions with fast electrons) before losing its momentum (mainly by
collisions with protons). The probability of its remaining in the
state $i$ is $r_{u,i}$ as the probability of recombination or further 
ionization is negligible.  Momentum absorbed during the transition in the
state $i$, $p_{lu}$, should then be redistributed to the two mass
reservoirs $mn_{i+1}$ and $mn_{i}$ of $i+1$ and $i$ ions in the
proportions 
$(1-r_{u,i})$ and $r_{u,i}$, respectively. Consequently, one gets:
\footnote{Note that these formulae
are valid for our radiative accelerations,
i.e. as defined by Eq. \ref{e4}.
Other formulae for the redistribution effect without 
$\frac{n_{i}}{n_{i+1}}$ factor would also hold but if combined
with a slightly different definition of radiative acceleration, e.g. that 
of \citet{Seaton97}.} 
\begin {equation}
a_{i} =  \frac{1}{mn_{i}} \sum_{lu} \frac{dp_{lu}}{dt} r_{u,i} =
\sum_{lu} a_{lu} r_{u,i}
\label{e8}  
\end {equation}
\begin {equation}
a_{i+1} =  \frac{1}{mn_{i+1}} \sum_{lu} \frac{dp_{lu}}{dt} (1-r_{u,i}) =
\sum_{lu} a_{lu} (1-r_{u,i}) \frac{n_{i}}{n_{i+1}}
\label{e9}  
\end {equation}
The redistribution function, $r_{u,i}$, depends on the ionization rate
from the particular level, $\beta_{u,i}$, and the collision rate of 
the ion, $\beta_{i}$:
\begin {equation}
r_{u,i} =\frac{\beta_{i}}{\beta_{u,i}+\beta_{i}}=
\frac{1}{\frac{\beta_{u,i}}{\beta_{i}}+1}
\label{e9b}  
\end {equation}

It is often useful to picture the radiative accelerations on many ions
being applied via some effective radiative acceleration on the element as
a whole.  Consequently, to compare the results with other authors 
and to display them, we also calculated the weighted mean 
radiative acceleration (although the results are stored
for each ion and abundance separately). We adopted the following
expression suggested by \citet{glam95}:
\begin {equation}
a(\tau) = \frac {\sum_{i} n_{i} D_{i} a_{i}}{\sum_{i} n_{i} D_{i}}
\label{e10}
\end {equation}
$D_{i}$ is obtained from $D_{i,\rm HI}, D_{i,\rm HII}$, which are 
diffusion coefficients which describe the diffusion of the $i$-th ion in
neutral and ionized hydrogen, respectively:
\begin {equation}
\frac {1}{D_{i}} = \frac {1}{D_{i,\rm HI}} + \frac {1}{D_{i,\rm HII}}
\label{e11}                                                
\end {equation}  
For the diffusion of the $i$-th ion (excluding neutral) in protons we
used \citep{ac60}:
\begin {equation}
D_{i,\rm{HII}} =\frac{1.947 \times 10^{9} T^{5/2} }
{n_{\rm{HII}}M_{2}Z_{i}^{2}A_{1}(2)}
\label {e12}
\end {equation}
where $n_{\rm{HII}}$ is the proton number density, $Z_{i}$ is the 
ionization degree
(0 = neutral, ...) and:
%
\[
A_{1}(2) =\ln (1+x_{\rm D}^{2});~~~~~~~M_{2}=\frac{A}{1+A}
\]
where $A$ is atomic weight in atomic mass units and:
%
\[
x_{\rm D} =\frac{4d_{\rm D}kT}{Z_{i}e^{2}}~~~{\rm and}~~~
d_{\rm D} =\sqrt \frac{kT}{4\pi n_{\rm e} e^{2}}~~~~~,
\]
where $d_{\rm D}$ is Debye length and $n_{\rm e}$ is electron number
density.
For the diffusion of neutrals in protons we used \citep*{mmr78}:
\begin {equation}
D_{0,\rm{HII}} =3.3\times 10^{4} \frac{T}{n_{\rm{HII}}} \sqrt
\frac{1+A}{\alpha A}
\label {e16}
\end {equation}
where $\alpha$ is the polarizability of neutral species
($\alpha =0.395, 0.667 \times 10^{-24}$ cm$^{3}$ for \nei\ and \hi\,
respectively, from \citet{tp71}). The same formula was used
for the diffusion of ions in neutral hydrogen with $\alpha
\rightarrow \alpha(\rm{H\,I})$ and $n_{\rm{HII}} \rightarrow
n_{\rm{HI}}$, where $n_{\rm{HI}}$ is the neutral hydrogen number
density. For the diffusion of neutrals in neutral hydrogen we used
\citep{ldv98}:
\begin {equation}
D_{0,\rm{HI}} = \frac{5.44\times 10^{3} \sqrt{T}}
{n_{\rm{HI}}(\delta_{\rm{HI}}+\delta_{i})^{2}} 
\sqrt \frac{1+A}{A}
\label {e17}
\end {equation}
where $\delta_{\rm{HI}}, \delta_{i}$ are \hi\ and diffusing atom 
diameters, respectively.
The following values calculated using the method of 
\citet*{crr63}\footnote{http://www.webelements.com/webelements.html} were
adopted:
$\delta_{\rm{HI}}=1.06\,10^{-8}\rm cm, \delta_{\rm{NeI}}=0.76\,10^{-8}\rm
cm$. 

\subsection{Line selection and numerical calculations}

\label{calculations} 

As one goes from the envelope higher into the atmosphere, the electron
density and associated Stark broadening drop sharply and the spectral
lines become much narrower. One needs to consider fine structure to
integrate properly the contributions from bound-bound transitions. Atomic
data for the Ne\,{\sc i-iv} lines were extracted from the Kurucz line list
\citep[][CDROM 23]{kuruczcd23}. However, one usually does not need to
consider all the lines of the element and it is of high practical
importance to select just the most important ones. Lines which are
important for a particular atmosphere model were selected in the following
way.

For practical purposes it is convenient to ascribe three 
characteristic Rosseland optical depths $\tau_{1}, \tau_{2}, \tau_{3}$ or 
characteristic temperatures 
$T_{\rm 1}(\tau_{1}) \le T_{\rm 2}(\tau_{2}) \le T_{\rm 3}(\tau_{3})$
to each ion $i$ of the element of interest. The range $T_{\rm 1}$ to 
$T_{\rm 3}$ spans the region where the ion $i$ is sufficiently populated
and its contribution to the total radiative acceleration of the element is 
thus not negligible, while $T_{\rm 2}$ is the approximate temperature 
at which the ion dominates the element and the $i$-th ion opacity is 
maximal.  For each line $lu$ of the ion $i$ two parameters $RA_{i,lu,1}$
and $RA_{i,lu,3}$ were calculated which correspond to two characteristic
temperatures $T_{\rm 1}, T_{\rm 3}$:
\[
RA_{i,lu,1,3} = 8.853 \times 10^{-13} m^{-1} \left[1-e^{- h\nu/(kT_{1,3})}
\right]
\]
\begin {equation}
f_{lu} g_{l,i} e^{- E_{l,i}/(kT_{1,3})} F_{\nu}(\tau _{\rm 1,3})
\label{e18}
\end {equation}
where $E_{l,i}$ is the excitation potential of the lower level and:
\begin {equation}
F_{\nu}(\tau_{\rm 1,3})=\pi B_{\nu}(T_{\rm eff})
~~~{\rm if}~~T_{\rm 1,3}<T_{\rm eff}, ~~~~~{\rm or} \\
\label{e19}
\end {equation}
\begin {equation}
F_{\nu}(\tau_{\rm 1,3})=\frac{4}{3}\frac{\pi}{\chi_{c,\nu}}
\frac{\partial B_{\nu}(T)}{\partial T}\frac{\partial T}{\partial r}
~~~{\rm if}~~T_{\rm 1,3}>T_{\rm eff}~~~,
\label{e20}
\end {equation}
where $B_{\nu}$ is the Planck function, $T_{\rm eff}$ is the effective
temperature, and $\chi_{{\rm c},\nu}$ the approximate continuous opacity
after \citet*{bmp79}.

The parameter $RA_{i,lu,1,3}$ is essentially a rough estimate of the
expected radiative acceleration through weak lines, as Eq.\ref{e20} takes
only opacity in the continuum into account and partition functions are set
to
1.\footnote{The first simplification does not have any serious effect on
this selection of the most important lines of a particular ion as it means
we overestimate the expected accelerations through strong lines, so we
cannot omit them. The second simplification also does not have any effect
on line selection of an ion as $RA_{i,lu,1,3}$ or radiative accelerations
for all the lines of a particular ion are scaled by the same value of ion
population which is proportional to its partition function. Again, it also
means we cannot omit important lines.} Then we calculate a sum
$RA_{i,1,3}= \sum_{lu} RA_{i,lu,1,3}$ through all the lines of the $i$-th
ion and skip the lines which do not contribute significantly to the sum.
We ended up with about 10$^3$ Ne lines for a particular model atmosphere.

To evaluate numerically the integral in Eq. \ref{e6} it is necessary 
to estimate the integration step and integration boundaries.
Thermal Doppler broadening at the coolest layer and classical radiative
damping put constraints on the width of any spectral line and we use 
minimum frequency spacings equal to 
\begin {equation}
\Delta \nu_{\rm step} \approx (\Delta \nu_{\rm
D}(\tau_{1})+\Gamma_{\rm R}/4\pi)/2
\label{e21a}
\end {equation}
where:
$\Delta \nu_{\rm D}=\nu c^{-1} \sqrt{2kT_{1}/m}$ and
$\Gamma_{\rm R}=2.47\times 10^{-22}\nu^{2}$.
Integration boundaries are associated with our line width, which
is generally the broadest at the place of highest electron density
due to Stark broadening and we use:
\begin{equation}
\Delta \nu_{\rm w}=k_{1}[\Gamma(\tau_{3})/4\pi+\Delta
\nu_{\rm D}(\tau_{3})]
\label{e21f}
\end{equation}
where
\begin{equation}
\Gamma(\tau_{3})=\Gamma_{\rm
R}+\Gamma_{\rm S}(\tau_{3})+\Gamma_{\rm W}(\tau_{3})
\label{e21d}
\end{equation}
is the sum of the radiative, Stark and Van der Waals damping parameters.
Here, $k_{1}$ is an adjustable constant for which numerical tests revealed
that $k_{1} \approx 2-3$ is high enough.  This is, however, just the case
of a weak line. If the opacity in the center of our line is much greater
than the opacity in the continuum one must usually integrate much further 
until the line opacity drops significantly below continuum. The reason why 
is apparent if one expresses the flux in Eq.\ref{e6} following the
diffusion approximation and neglects continuum opacity; it is inversely 
proportional to the Voigt function. Consequently,
the integrand in Eq.\ref{e6} is a constant function of frequency.
Only when the integration is carried far enough into the line wing,
where continuum opacity again dominates, does flux approach a constant
function of frequency and the integrand sharply fall.
Finally, the integration boundaries should be the maximum value of 
the mentioned effects:
\begin {equation} 
\Delta \nu_{\rm width}=\pm \rm MAX [\Delta \nu_{w}(\tau_{3}); 
\Delta \nu_{\rm s}(\tau_{2})]
\label{e21b}
\end {equation}
where $\Delta \nu_{s}(\tau_{2})$ 
is defined for strong lines only and is $\Delta \nu$ for which the
line opacity drops significantly below the continuum opacity at the depth
where the involved ion population reaches its peak values.
Typically, anything from 40 to 10$^3$ frequency points were
necessary to calculate the radiative acceleration through a Ne line.

\begin{figure}
\epsfig{file=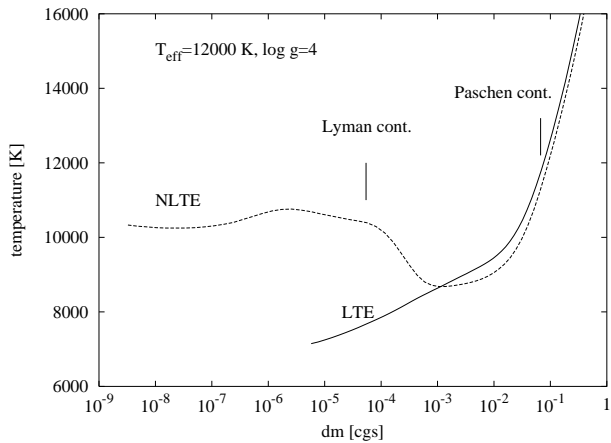,width=8.5cm}
\caption{Temperature behaviour in a representative Kurucz's line blanketed
LTE and Hubeny's NLTE model atmosphere.
Vertical bars indicate the unit monochromatic optical depth in the Lyman
and Paschen continua close to the important \nei\ $\lambda$736 and 
\nei\ $\lambda$6402 transitions. As a depth coordinate we use `dm'
(g cm$^{-2}$) which is the mass of a column above a unit area at that
depth in the atmosphere.}
\label{f1}
\end{figure}

The models adopted here were NLTE model atmospheres calculated with
{\sc tlusty195} \citep{hubeny88,hl92,hl95} for standard solar composition
and zero microturbulence. We treated \hi, He\,{\sc i}, He\,{\sc ii},
C\,{\sc i}, C\,{\sc ii}, C\,{\sc iii} as explicit ions, which means that
their level populations were solved in NLTE under the assumption of
statistical equilibrium and their bound-bound, bound-free, and free-free
opacities were taken into account. Carbon was included as it is an
important opacity source at cooler effective temperatures
\citep{hubeny81}. In Fig. \ref{f1} we plot the NLTE model against the
\citet{kuruczcd13} LTE line blanketed model for the same effective
temperature and gravity. The two vertical bars point to the area where the
continua around \nei\ $\lambda$736 and \nei\ $\lambda$6402 lines are
formed. The first is a strong resonance line in the Lyman continuum and
the second is in the Paschen continuum in the visible. Abundance analyses
of Ne have been performed using $\lambda$6402 \citep{db00}. The Lyman
continuum is very opaque and models which go very high in the atmosphere
are needed to solve properly the radiative transfer at these wavelengths.

$F_{\nu}(\tau)$ in Eq. \ref{e6} was calculated using the {\sc{synspec36}}
and {\sc{synspec42}} \citep*{hlj94} codes to solve the
radiative transfer modified to get a nonstandard output of flux at each
frequency and depth (I. Hubeny, priv. comm.), and to take
into account the elements with atomic number $>30$ \citep{Krticka98}. This
enabled us to take into account the bound-free and free-free opacity from
explicit levels specified in the atmosphere model and the line blending
using the full line list of \citet{kuruczcd23}. All the Kurucz lines with
line-to-continuum opacity ratio greater than $10^{-8}$ at $\tau={\rm
MAX}[1,\tau_{2}]$ were considered. All the resonance lines of Ne and other
elements were calculated in approximate NLTE (allowing the source
function to deviate from $B_{\nu}$ following the `second-order escape
probability method' of \citet{Rybicki84}, see Eq. \ref{e23}).

Radiative accelerations were then calculated following Eq. \ref{e6} 
assuming $b_{u,i}=b_{l,i}=1$ and LTE level populations.  Partition
functions were taken from the {\sc{uclsyn}} code
\citep{sd88,smith92}.  
Radiative accelerations were calculated for
Ne\,{\sc i-iv} and four abundances: $50\times A_{\sun}, 
A_{\sun}, 10^{-2} \times A_{\sun}, 10^{-5} A_{\sun}$
where $A_{\sun}=1.23 \times 10^{-4}$
and for the following model atmosphere parameters:
$T_{\rm eff}=~11\,000, 12\,000, 13\,000, 14\,000, 15\,000$\,K and $\log
g=4$, plus models with $T_{\rm eff}=12\,000$\,K and $\log g=3.5, 4.5$.
These are available in digital form  
in \citet{bds02} including a {\sc{fortran77}} code containing 
the partition function routines.


\section{Results and discussion}

\subsection{Radiative accelerations on different Ne ions}

\begin{figure}
\epsfig{file=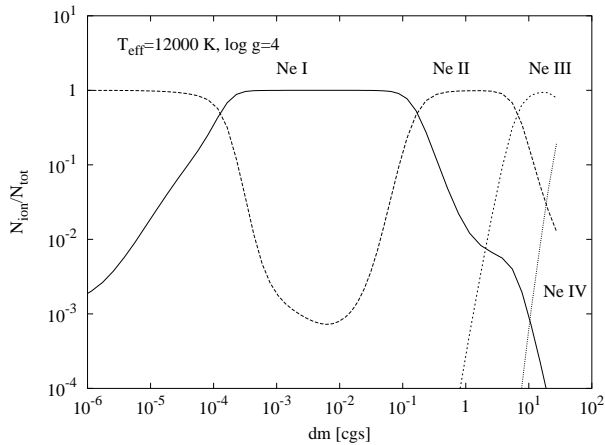,width=8.5cm}
\caption{Ne ionization fractions in LTE in a representative
NLTE model atmosphere.}
\label{f2}
\end{figure}

The LTE populations of various Ne ions throughout the atmosphere are
depicted in Fig. \ref{f2} for a representative atmosphere model with
$T_{\rm eff}=12\,000$\,K, $\log g=4$. \nei\ is definitely dominant in the
atmosphere.  Ne\,{\sc iv} becomes populated at the base of our models
while \neii\ may dominate at very low densities and electron
concentrations for $dm<10^{-4}$ g\,cm$^{-2}$. But above
$dm<10^{-3}$g\,cm$^{-2}$ strong departures of Ne ionization and
excitation equilibrium from LTE occur \citep{sigut99} and our
accelerations on different Ne ions should be considered as approximate
above this level (see Fig.\ref{f8} for an illustration of expected
deviations from fully consistent NLTE case).

\begin{figure}
\epsfig{file=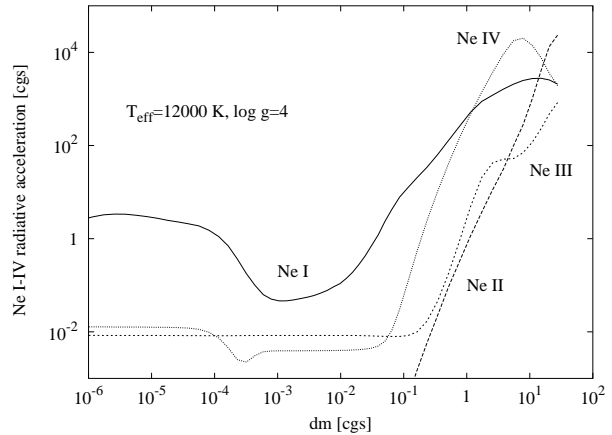,width=8.5cm}
\caption{Radiative accelerations on Ne\,{\sc i-iv} ions in a
representative
NLTE model atmosphere for the standard abundance of Ne. 
The downward gravitational acceleration of $10^4$ cm\,s$^{-2}$ is at least 
2-3 orders of magnitude greater than the upward radiative acceleration 
throughout the photosphere.}
\label{f3}
\end{figure}

Accelerations in the representative model are plotted in Fig. \ref{f3}.
One can see that none of them exceeds the gravitational acceleration
except at the very deepest layers. The acceleration on \nei\ is strongest
in the atmosphere where this ion is also most populated. The
contribution of the neutral species like \nei\ to the total effective
radiative acceleration of the element is both favoured by weighting with 
their diffusion coefficient (see Eq. \ref{e10}) which is approximately two
orders of magnitude larger than that for the charged species and reduced
by the redistribution effect. Nevertheless, acceleration
of \nei\ will be the most important contribution to the total
acceleration not only in the atmosphere but also at rather great depth up
to $dm \simeq 10$\,g\,cm$^{-2}$. The low radiative accelerations found on
\nei\ are a consequence of its atomic structure because all resonance
transitions, which are usually the most important, are in the Lyman
continuum where \nei\ atoms see little photon flux, while the rest of the
lines originate from highly excited and weakly populated states. Nor can
the redistribution effect discussed earlier increase the Ne I acceleration
(see Eq.~\ref{e8} and Fig.~\ref{f6b}); just the opposite occurs. 
Radiative acceleration
gained in the \nei\ state is redistributed between \nei\ and {\neii},
which reduces the \nei\ acceleration and adds it to the \neii\ state
where, however, the acceleration is not so effective due to the much lower
diffusion coefficient of ionized species; see Eqs. \ref{e7}, \ref{e10}.

We would like to stress that radiative acceleration (Eq. \ref{e6}) on a
particular ion is very sensitive to the partition functions. These are
generally precise in the region where the ion dominates the element but
have large uncertainties outside
the region. Displaying effective accelerations overcomes the problem as
partition functions drop out in Eq. \ref{e10} outside the region.
Thus, if the reader wishes to use our radiative accelerations stored
separately for each ion, the same partition functions should be used also.

\subsection{Dependence on effective temperature and surface gravity}

\begin{figure}
\epsfig{file=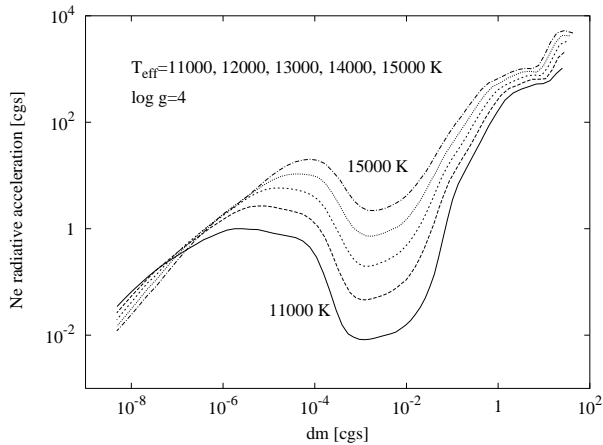,width=8.5cm}
\caption{Radiative accelerations on Ne throughout the atmosphere 
as a function of stellar effective temperature and depth.
Calculated for standard abundance of Ne.}
\label{f4}
\end{figure}

Now we can proceed further and explore how the situation changes with the
effective temperature of the star. It is clear from Fig. \ref{f4} that
the total acceleration rapidly increases with the effective temperature.
However, within our range of interest it never exceeds the gravity in the
line forming region. Nevertheless, this suggests that Ne could accumulate
at deeper layers in cooler stars than in hotter ones, the latter
thus being more vulnerable to departures from the ideal stable atmosphere.
Various mixing processes or radiatively driven stellar winds whose
intensity also increases with the effective temperature
\citep{babel95,kk01} might more easily enrich Ne in the
atmospheres of much hotter stars. This resembles the case of He which has 
atomic structure similar to Ne. The abundance of He progressively
increases with effective temperature from He-weak towards He-rich stars.
This suggests that it might be interesting to search for Ne-rich
analogues of He-rich stars among early B stars. On the other hand, for
$\teff<11\,000$\,K, hydrogen becomes partially ionized, which triggers
ambipolar diffusion and hydrogen superficial convective zones. This may
also transport some Ne to the atmosphere from below. Our observations
\citep{db00} indicate that Ne underabundances are most
pronounced in the middle of the HgMn domain and tend to be less extreme
towards the cool and hot ends of the HgMn temperature region. This
pattern could therefore be qualitatively expected. Observing Ne below
$\teff = 11\,000$\,K, however, becomes very difficult.

\begin{figure}
\epsfig{file=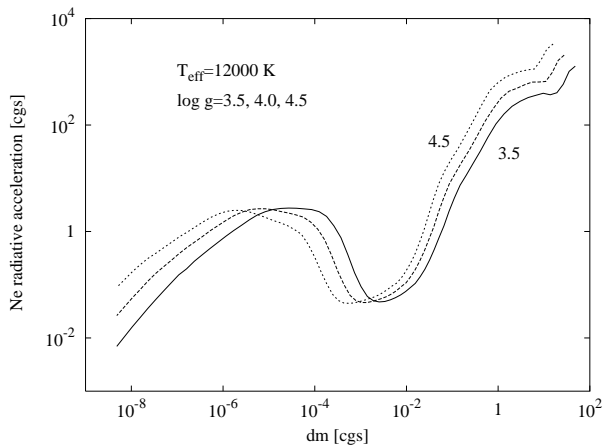,width=8.5cm}
\caption{Radiative accelerations on Ne throughout the atmosphere 
as a function of stellar surface gravity and depth.
Calculated for standard abundance of Ne.}
\label{f5}
\end{figure}

The radiative accelerations also depend on the surface gravity (Fig.
\ref{f5}). Lowering the surface gravity of the model shifts the radiative
acceleration behaviour to deeper layers and reduces the radiative
acceleration at the bottom of the atmosphere. However, the gravity to
radiative acceleration ratio does not change very much and in this context
effective temperature is a more important atmospheric parameter than $\log
g$. Apparently, within the main sequence (approximately $3.5\leq \log g
\leq 4.5$), radiative acceleration of Ne never overcomes gravity.

\subsection{Effects of homogeneous Ne abundance}

\begin{figure}
\epsfig{file=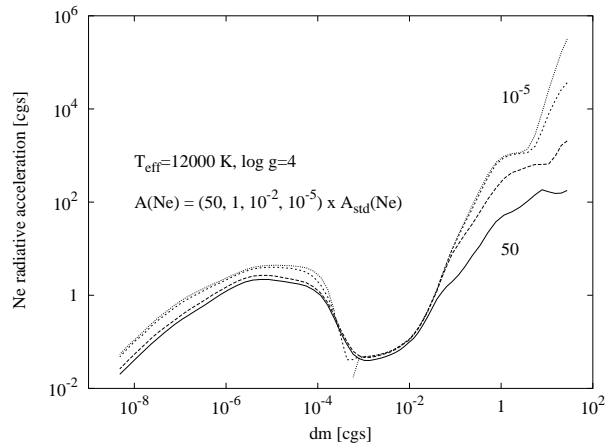,width=8.5cm}
\caption{Radiative accelerations on Ne throughout the representative
NLTE model atmosphere as a function of the homogeneous Ne abundance.}
\label{f6}
\end{figure}

Radiative acceleration on the element also depends on its abundance. This
is a simple consequence of the fact that the energy flux in the integral
in Eq. \ref{e6} depends on the element abundance.  This is, however, only
important in the case of strong lines where the line opacity is comparable
to or larger than opacity in the continuum. One can then expect that
radiative acceleration will asymptotically increase as element abundance
decreases, until it reaches some maximum value corresponding to the
sufficiently low abundance when all the element lines disappear and
radiative acceleration will no longer be sensitive to the abundance
\citep[for more detail see][]{al00}.  The situation is illustrated in Fig.
\ref{f6}. Surprisingly, the accelerations are very little dependent on
abundance in the photospheres above $dm \simeq 0.1$ g\,cm$^{-2}$ and do
not exceed gravity for any abundance except at the extreme base of our
models. Here a Ne deficit ranging from -0.5 dex for the hottest to -2 dex
for the coolest model could be supported by radiation (i.e. radiative
acceleration roughly equals gravity for this abundance and depth,
$dm\approx 20$\,g\,cm$^{-2}$).  This weak abundance sensitivity results
from the fact that resonance \nei\ lines are not very strong because they
are in the Lyman continuum where the continuous opacity is very large,
effectively reducing line to opacity ratio. We have found that
contributions to the radiative acceleration of lines originating from
excited \nei\ levels is also important, but these weak
lines do not introduce strong dependence on the abundance because their
line-to-continuum opacity ratio is low due to the high excitation energies
and low populations of the levels from which they originate. One can see
from the figure that there is no abundance for which the radiative
acceleration could balance the gravity in the line forming region above
$dm \simeq 0.1$\,g\,cm$^{-2}$. The implication is that in this region,
assuming a stable atmosphere devoid of any motions, Ne should sink and
should be almost completely depleted and only its strong concentration
gradient could balance the downward flux of Ne atoms. The Ne atom is at
least an order of magnitude heavier than the mean `molecular' mass of the
gas, so the characteristic scale of the Ne abundance gradient 
would be an order of
magnitude less than the pressure scale height, i.e.  $\log dm \approx
0.1$. Consequently, the fact that Ne is detected in most HgMn stars
\citep{db00} suggests the presence of some kind of instability competing
with diffusion and suppressing the efficiency of diffusion processes. An
example of such an instability is the weak stellar wind suggested by
\citet{ldv98}.

One may also notice a sudden drop in the radiative accelerations at
approximately $10^{-4} \leq dm \leq 10^{-3}$ g\,cm$^{-2}$. This is not a
numerical artifact but a very interesting effect when radiative
acceleration may acquire negative values; this will be discussed in more
detail below.

\subsection{Redistribution effect}

\begin{figure}
\epsfig{file=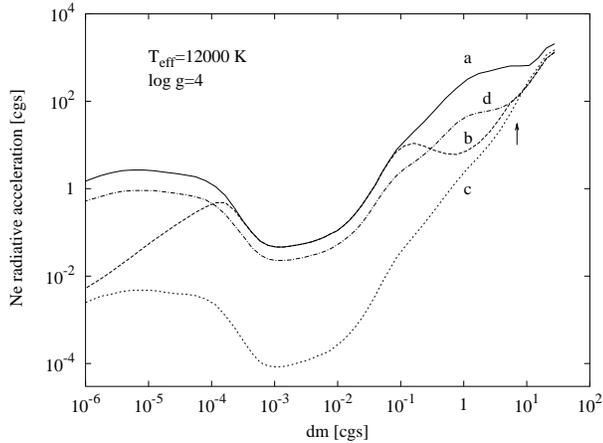,width=8.5cm}
\caption{An illustration of the redistribution effect on the 
radiative accelerations on Ne throughout the representative
NLTE model atmosphere as a function of depth calculated for standard 
homogeneous abundance of Ne. 
(a) -- no redistribution -- solid line,
(b) -- no redistribution and no diffusion coefficient weighting -- dashes,
(c) -- \citet{glam95} type redistribution -- dots, 
(d) -- our type redistribution -- dash-dots. 
The arrow points to the depth where $T=31000K$.}
\label{f6b}
\end{figure}

To take the redistribution effect into account properly one would need
to calculate probabilities of all possible radiative and collisional
transitions from the upper level to continuum including multiple
cascade transitions as
suggested by \citet{glam95}. That is a very difficult task and 
the above authors developed a method in which they set $r_{u,i}=0$ for
$n \geq 3$ and $r_{u,i}=1$ for lower states. Thus their $r_{u,i}$
is a simple step function of level energy and does not depend on 
e.g. temperature.
To cope with the problem and estimate the influence of
the redistribution effect we propose a slightly different and
more general approach. The idea is simple. 
As mentioned in Section~\ref{theory}, collisions with electrons may
change the ionization state before the momentum absorbed by the atom
is lost in collisions with protons. Consequently, the redistribution 
function from a particular excited level depends mainly on 
the electron to proton collisional rates ratio.
Higher atomic levels with
energies corresponding to the electron kinetic energy $\approx kT$ below
the ionization threshold are most strongly coupled to the continuum 
via collisional excitation and ionization.
Probabilities of such processes are strongly 
temperature dependent via the Boltzmann factor $e^{-\Delta E/kT}$  
\citep{Seaton62,regemorter62}
and still deeper atomic levels are affected as one goes deeper
into the star. Electron to proton number density ratio almost 
does not change with depth in such hot stars and the Boltzmann factor 
thus embraces the essence of the electron to proton collision rates ratio
and one may write $\beta_{u,i}/\beta_{i}=C e^{-\Delta E/BkT}$.
Consequently, based on Eq.~\ref{e9b} we suggest the following expression 
for the $r_{u,i}$:
\begin {equation}
r_{u,i}=\frac{1}{Ce^{-U}+1}~~ {\rm with}~~
U=(I_{i}-E_{u,i})\left(\frac{1}{A}+\frac{1}{BkT}\right)
\label{e21c}
\end {equation}
where $I_{i}$ is the $i$-th ion ionization potential and $A,B,C$ are
adjustable constants. It is beyond the scope of the present paper
to calibrate precise values for the above mentioned constants. 
$B$ is most important as it parameterises the temperature dependence and 
is of the order of 1. The $1/A$ term was introduced 
for those who prefer to use a \citet{glam95} type of redistribution 
but do not like its step function shape and $A$ is of the order
of $kT$, but we use $A\rightarrow \infty$. 
$C$ parameterises the ratio of the total
electron to proton collision rates. If $E_{u,i} \rightarrow I_{i}$ 
the level will register all electron impacts and $\beta_{u,i}/\beta_{i}=C$.
Because electrons are $\sqrt{m_{p}/m_{e}}$ faster than protons, their
collisions will be more frequent 
by approximately the same factor, and C is thus of the order of 40.
The method of \citet{glam95} is equivalent to the approximation of 
Eq.\ref{e21c} by a temperature independent rectangle and could be 
approached as a special case
when $B\rightarrow \infty$ and $A\rightarrow 0$ or $A\rightarrow \infty$ 
depending on whether
$E_{u,i}>I_{i}-E_{crit}$ or $E_{u,i}<I_{i}-E_{crit}$, respectively, where
$E_{crit}$ is some threshold energy. Note that $A\rightarrow 0$ 
means $r_{u,i}\rightarrow 1$ i.e. no redistribution at all and 
$A\rightarrow \infty$ means $r_{u,i}\rightarrow 1/(C+1)\rightarrow$ a 
small value, i.e. total redistribution is approached but never fully 
realised.  Assuming that \citet{glam95} took the lowest $n=3$ level of 
C\,{\sc ii} 
as a cutoff in their calculation for $T=31000$K which corresponds to 
$E_{crit}=80000$ cm$^{-1}$ below continuum, we can simulate their method
(using our Eq.~\ref{e8},~\ref{e9} and forcing $r_{u,i}=1$ or $r_{u,i}=0$)
and calibrate our method so that it
would give $r_{u,i}=0.5$ at the same temperature. Assuming 
$C=40, A \rightarrow \infty$ we get from Eq.~\ref{e21c} that $B=1.0$ 
-- a reasonable value indeed. 
Results of both methods are illustrated in Fig. \ref{f6b} and one can
see that while both methods may give similar results for $T=31000$K
the \citet{glam95} method strongly overestimates the redistribution effect
for lower temperatures.
A simple test of the redistribution effect treatment is
recommended by calculating radiative accelerations
without weighting them by diffusion coefficients (setting $D_{i}=1$ for
example in Eq.~\ref{e10}). This is because such accelerations are
an invariant of the redistribution process.

\subsection{Comparison with other work}

\begin{figure}
\epsfig{file=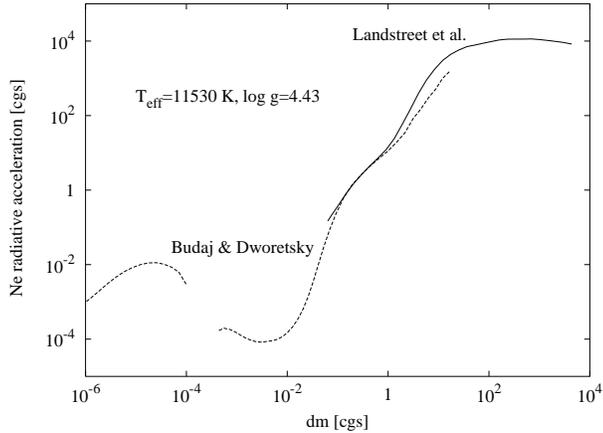,width=8.5cm}
\caption{Comparison of our calculations in the atmosphere with those of
\citet{ldv98} in the envelope for the same model parameters. Here, for
\nei\, only resonance lines are considered. Both were calculated for
standard abundances of Ne.}
\label{f7}
\end{figure}

We calculated an atmosphere model for $T_{\rm eff}=11\,530$\,K, $\log
g=4.43$, corresponding to one of the envelope models of \citet{ldv98}.
Then we calculated the radiative accelerations, but in the case of \nei\
we omitted all except the resonance lines, to simulate the assumption
adopted by \citeauthor{ldv98}.  This assumption is not really
justified as the omitted lines contribute significantly to the radiative
acceleration. Inclusion of \nei\ lines originating from excited levels,
which absorb at the wavelengths where a star radiates much more than in
the Lyman continuum, may increase the acceleration on \nei\ by more than
1.5 dex. However, this does not have any serious consequences on the
conclusions of \citeauthor{ldv98}, as the radiative accelerations are so
small they will still remain below the gravity. The comparison is shown in
Fig. \ref{f7}. In this case we also considered $D_{i}=1$ in Eq. \ref{e10}
to follow their method. The agreement is quite good and small departures
at the base of the atmosphere are mainly caused by slightly different
temperature behaviour between our model atmosphere and their envelope
model.  The curious gap in our curve at $10^{-4} \leq dm \leq
10^{-3}$\,g\,cm$^{-2}$ appears because the acceleration is negative here
and cannot be plotted on a logarithmic scale.

\subsection{NLTE effects}

The aforementioned negative acceleration is a very interesting NLTE
effect which may lower still further the radiative accelerations on some
ions in this part of the atmosphere. It is a consequence of the
temperature inversion in the model atmosphere and the atomic structure of
\nei. As shown in Fig. \ref{f1}, temperatures in the outer layers rise in
a NLTE
model for $10^{-4} \leq dm \leq 10^{-3}$\,g\,cm$^{-2}$. At this depth, and
in these stars, the atmosphere is still optically thick in the Lyman
continuum.  Thus, according to the diffusion approximation, the star
radiates into itself towards cooler regions at these wavelengths. If an
element happens to have its important transitions in the Lyman continuum
it may be pushed into the star by the radiation and its acceleration is
thus negative.  Similar effects can be expected also in the cores of very
strong lines not necessarily in the Lyman continuum.
To explore NLTE effects on the results we also calculated
radiative accelerations on \nei\ for three different assumptions in a
representative NLTE model atmosphere with $T_{\rm eff}=15\,000\,{\rm K},
\log g=4$ and zero microturbulence:
\begin{enumerate} 
\item LTE case: neon populations in Eq. \ref{e6} are in LTE and 
the neon lines' source functions for flux
calculations is equal to the Planck function
\begin {equation}
S_{\nu}=B_{\nu};
\label{e22}
\end {equation}
\item approximate NLTE: the same, 
but the line source function for resonance transitions is allowed
to deviate from the Planck function following second order escape
probability methods 
\begin {equation}
S_{\nu}=\sqrt{\frac{\epsilon}{\epsilon+(1-\epsilon)K_{2}(\tau)}}B_{\nu}
\label{e23}
\end {equation}
where $\epsilon$ is the photon destruction probability and
$K_{2}(\tau)$ is the kernel function
\citep[see][the {\sc{synspec}} manual, for more detail]{hlj94}. 
Note that this option was used in the calculations
described in Section~\ref{calculations} above; 
\item fully consistent NLTE case: 
with NLTE model atmosphere, NLTE populations and NLTE source functions
\begin {equation}
S_{\nu}=\frac{2h\nu^{3}}{c^{2}}
\left(\frac{g_{u,i}n_{l,i}}{g_{l,i}n_{u,i}}-1\right)^{-1}.
\label{e24}
\end {equation}
In the full NLTE case the
NLTE model atmosphere and \nei\ level populations were calculated 
with {\sc tlusty195}. As an input for {\sc tlusty195} we considered
Hubeny's \hi\ atom model with 9 explicit levels and continuum, the 31
level \nei\ atom model taken from \citet{db00}, and a simple
4 level atom of \neii\ with continuum (Table~\ref{t1}). 
Generally, the data for Ne bound-bound and bound-free transitions 
were taken from the TOPbase \citep{cmoz93,hs94} but \nei\ oscillator 
strengths were from \citet{Seaton98}. The
{\sc{modion}} IDL interface written by \citet{vldhh95} was
particularly useful in constructing \nei\ and \neii\ atom models.
Populations of those very high \nei\ levels not included in our \nei\ atom
model were considered in LTE relative to the ground state of {\neii},
of which the population was calculated in NLTE, when calculating the
energy flux and acceleration.
All \nei\ level populations were set to LTE below $dm>0.3$\,g\,cm$^{-2}$
when calculating the radiative accelerations.
\end{enumerate}

\begin{table}
\caption{\neii\ energy levels considered.
Column 1: level designation,
Column 2: ionization energy in cm$^{-1}$,
Column 3: statistical weight of the level. Energies are from \citet{persson}.}
\begin{center}
\begin{tabular}{lll}
\hline
Desig. &Energy    & g  \\
\hline
$2p^5~^2\!P^o$ & 330445. &     6 \\
$2p^6~^2\!S$   & 113658. &     2 \\
$3s~^4\!P$     & 111266. &    12 \\
$3s~^2\!P$     & 106414. &     6 \\
\hline 
\end{tabular}
\end{center}
\label{t1}
\end{table} 

\begin{figure}
\epsfig{file=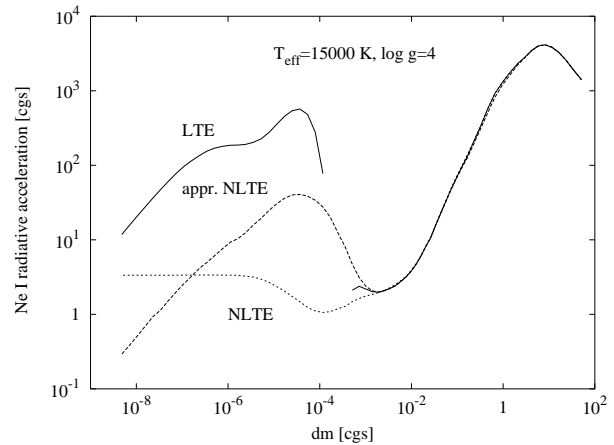,width=8.5cm}
\caption{Comparison of radiative accelerations on \nei\ obtained
using different assumptions: LTE, approximate NLTE and full
NLTE (see text). Calculated for standard abundance of Ne.} 
\label{f8}
\end{figure}

The differences between the three assumptions are exhibited in Fig.
\ref{f8} and become crucial for $dm<10^{-3}$\,g\,cm$^{-2}$, where the
radiative accelerations in LTE and NLTE may differ by more than 3 dex.  
In the case of LTE we observe negative acceleration in the region of the
temperature inversion. This was explained above. The approximate NLTE case
is parallel to the LTE acceleration but about 1.5 dex smaller. This is
caused by the difference in computed energy fluxes. While flux in the
approximate NLTE case is real - lines are in absorption - and surprisingly
close to the flux in the full NLTE case, the resonance \nei\ lines are in
emission in LTE as soon as the atmosphere becomes optically thin in the
Lyman continuum and acceleration soars immediately. The difference between
approximate NLTE and full NLTE originates mainly in the differences in
level populations.

\section{Summary} 

We have calculated radiative accelerations on Ne\,{\sc i-iv}
ions in the atmospheres of late B main sequence stars. We take into
account
the fine structure and calculations include the
effects of line blending using the whole Kurucz line list, bound-free and
free-free opacity of explicit elements (H, He, C) as well as some NLTE
effects.  We explored how the acceleration changes with respect to
effective temperature, surface gravity and homogeneous Ne abundance and
found that it is much smaller than the gravity in and above the observed
line
forming region. Only at the base of our models ($dm \simeq
20$\,g\,cm$^{-2}$) 
could a Ne deficit of about 0.5-2.0 dex be supported by
the radiation, depending on the effective temperature. This implies that,
in the stable atmospheres of late B stars, Ne should be almost completely
depleted from the photosphere. This is qualitatively in agreement with
the observations of \citet{db00} but the fact that we
detected Ne in some HgMn stars suggests the presence of some weak
transport or mixing mechanism contaminating observed layers by Ne from the
reservoir underneath. Possibly, this mechanism is associated with the
partial hydrogen ionization at the cool end and a very weak stellar wind
at the hot end of the HgMn \teff\ span.

Finally, we demonstrated by a full NLTE calculation the importance
of such effects on radiative accelerations and slightly improved a current
treatment of the redistribution effect which should be used in future
calculations.

\section*{Acknowledgments}

We thank I. Hubeny and J. Krti\v{c}ka for patient
discussions and help with {\sc{tlusty}} and {\sc{synspec}} codes, B.
Smalley for his work on fitting partition functions for {\sc{uclsyn}},
J.D. Landstreet and G. Alecian for discussing details about their
calculations and an anonymous referee for many useful comments.  
J.B. gratefully acknowledges the support of the Royal
Society/NATO fellowship scheme (Ref. 98B) and partial support by the VEGA
grant No. 7107 from the Slovak Academy of Sciences and APVT-51-000802
project.

\end{document}